\documentclass[11pt,a4paper]{article}

\usepackage[paperwidth=25cm]{geometry}
\usepackage[T1]{fontenc}
\usepackage[latin1]{inputenc}
\usepackage{graphicx}
\usepackage{amsmath}
\usepackage{amssymb}

\newtheorem{theorem}{Theorem}

\newcommand{\bl}{\boldsymbol}

\newcommand{\lef}{\left(\,}
\newcommand{\rig}{\,\right)}
\newcommand{\ph}{\phantom}

\newcommand{\eq}{\,=\,}
\newcommand{\ma}{\,+\,}
\newcommand{\me}{\,-\,}

\begin{document}

\title{\textbf{Integrability Conditions for Killing-Yano Tensors and Conformal Killing-Yano Tensors}}

\author{\textbf{Carlos  Batista}\\
\small{Departamento de F\'{\i}sica}\\
\small{Universidade Federal de Pernambuco}\\
\small{50670-901 Recife-PE, Brazil}\\
\small{carlosbatistas@df.ufpe.br}}
\date{}




\maketitle

\begin{abstract}
The integrability conditions for the existence of a conformal Killing-Yano tensor of arbitrary order are worked out in all dimensions and expressed in terms of the Weyl tensor. As a consequence, the integrability conditions for the existence of a Killing-Yano tensor are also obtained. By means of such conditions, it is shown that in certain Einstein spaces one can use a conformal Killing-Yano tensor of order $p$ to generate a Killing-Yano tensor of order $(p-1)$. Finally, it is proved that in maximally symmetric spaces the covariant derivative of a Killing-Yano tensor is a closed conformal Killing-Yano tensor and that every conformal Killing-Yano tensor is uniquely decomposed as the sum of a Killing-Yano tensor and a closed conformal Killing-Yano tensor. \textsl{(Keywords: Conformal Killing-Yano tensors, Conformal Killing tensors, Integrability conditions, General relativity, Maximally symmetric spaces)}
\end{abstract}

\section{Introduction}

The so-called hidden symmetries of curved manifolds, represented by Killing tensors and Killing-Yano (KY) tensors, have proved to be invaluable tools to the development of General Relativity, both from the Physical and Mathematical points of view. They yield conservation laws that enable the separability of differential equations, which, in turn, can lead to the integrability of equations of motion. For instance, the geodesic equation in $4$-dimensional Kerr spacetime could be fully integrated thanks to the existence of a non-trivial Killing tensor of order two \cite{Carter-constant,Walk-Pen}. Moreover, this Killing tensor have enabled the integration of Klein-Gordon and Dirac field equations in Kerr background \cite{Carter-KleinG,Chandra-Dirac} as well as the separability of gravitational and electromagnetic perturbations \cite{Teukolsky}. Since such Killing tensor turns out to be the square of a Killing-Yano tensor of order two \cite{Collinson,Steph_KY}, we can say that these integrability properties are due to the existence of a KY tensor in Kerr spacetime. Analogously, conformal Killing-Yano tensors (CKY) are associated with conservation laws along null geodesics and integrability of massless field equations. Furthermore, in \cite{JezierskyAF} the CKY tensors were used to motivate a suitable definition of asymptotic flatness in 4-dimensional spacetimes.

More recently, Killing-Yano tensors proved to be of great relevance in higher-dimensional spacetimes as well. In Ref. \cite{Kastor} these tensors were used to define gravitational charges in spacetimes that are asymptotically flat in a restricted number of spatial directions, which is of applicability in manifolds with branes. Additionally, it was shown that the family of Kerr-NUT-(A)dS spacetimes \cite{KerrNutAds} admits, in arbitrary dimension, a set of Killing-Yano tensors of various orders \cite{Frolov_KY}. These KY tensors were then used to integrate the geodesic equation \cite{Kubiz,Krtous} as well as the Klein-Gordon \cite{Frol-KG} and Dirac equations \cite{Oota} in such background. The issue of separability of gravitational perturbations was investigated in \cite{OotaPerturb}. Interestingly, it turns out that all these KY tensors necessary to achieve integrability in Kerr-NUT-(A)dS spacetimes can be constructed from a single conformal Killing-Yano tensor that is closed \cite{Frolov_KY}. For a review on the role played by CKY tensors in exact solutions of Einstein's equation see \cite{Yasui}. It has also been demonstrated a connection between the existence of a closed CKY tensor and the integrability of maximally isotropic distributions \cite{Mason_KY}, \textit{i.e.}, distributions generated by pure spinors. These facts make evident the huge importance of CKY tensors to higher-dimensional gravitational theories.

The intent of the present article is to investigate the integrability conditions for the existence of a conformal Killing-Yano tensor of arbitrary order in a manifold of any dimension. An approach toward the investigation of the same problem have also been done by Kashiwada \cite{Kashiwada-Int} \footnote{I thank Tsuyoshi Houri for pointing out this reference. Unfortunately, my attention have been driven to Kashiwada's work only after I released the preprint of the present article, so that this article have a considerable overlap with \cite{Kashiwada-Int}. The main reason for such overlap is that both works share the same goal, namely generalize Tachibana's results obtained in \cite{Tachibana-CKY} to CKY tensors of arbitrary rank. In spite of the similarities, the present article adds some new contributions, as the recognition that the integrability condition can be expressed just in terms of the Weyl tensor and the analysis of some of its consequences. Moreover, here all steps on the deduction of the integrability condition are explicit, whereas in \cite{Kashiwada-Int} some cumbersome manipulations are omitted. Finally, the procedure adopted here is simpler, since just few indices are permuted during the manipulations, while in \cite{Kashiwada-Int} all free indices are involved in the permutations.}. The particular case of KY and CKY tensors of order two was already done by Tachibana in \cite{Tachibana-KY} and \cite{Tachibana-CKY} respectively, while Killing tensors of order two were studied in \cite{Hauser}. Killing-Yano tensors of order $n-1$, with $n$ standing for the dimension of the manifold, were fully analysed in \cite{Bat-KYn-1}. Furthermore, the integrability conditions for Killing spinors in 4-dimensional spacetimes have been addressed  in \cite{KillingSpinors}. Such integrability conditions proved to be valuable in 4-dimensional General Relativity, since they have provided connections between the algebraic type of the Weyl tensor and the existence of hidden symmetries \cite{Walk-Pen,Collinson,Hugh_Killing}. Hopefully, the study performed here will, likewise, be of relevance to higher-dimensional General Relativity.

In addition to General Relativity, Killing-Yano tensors and its conformal relatives have also applicability in other areas. For instance, they can be used to define symmetry operators that commute with the D'Alembertian and Dirac operators \cite{Benn-DiracSymme}, which is of relevance to quantum field theory. Notably, while classical symmetries associated to KY tensors are preserved at the quantum level, those associated to Killing tensors generally are not \cite{Santillan}.
One can also use KY tensors to find conserved charges and integrate equations of motion of conservative systems in classical mechanics \cite{KY-ClassMech}, as well as to solve Maxwell's equation in curved spaces \cite{DebyePot}. Killing-Yano tensors have further been used to construct Lax pairs in curved manifolds \cite{KY-Lax}, which is of interest for the theory of integrable systems. Moreover, hidden symmetries proved to be of relevance in the study of supersymmetric systems \cite{KY-SUSY,Santillan}.

The outline of the present article is as follows. Section \ref{Sec.Definitions} sets the notation and reviews the basic definitions concerning conformal Killing-Yano tensors. In Sec. \ref{Sec.IntegrabCond} the integrability conditions for the existence of a CKY tensor of arbitrary order in arbitrary dimensions are worked out. As a bonus, the integrability conditions for the existence of KY tensors as well as closed CKY tensors are also obtained. Particularly, the
allowed algebraic types for the Weyl tensor in a 4-dimensional manifold admitting a CKY tensor of order two are displayed in Sec. \ref{Sub.Sec.p2}. Then, in Sec. \ref{Sec.EinsteinSpaces} it is shown how one can use a CKY tensor of order $p$ to construct a KY tensor of order $(p-1)$ in Einstein spaces with constrained Weyl tensors. The particular case of $p=3$ in a 4-dimensional Lorentzian manifold is explicitly worked out as an example. Finally, in Sec. \ref{Sec.MaximallySym} some results regarding conformal Killing-Yano tensors in maximally symmetric spaces are proved. Particularly, it is shown that every CKY tensor in a maximally symmetric space can be decomposed as the sum of a KY tensor and a closed CKY tensor.

\section{General Aspects of Conformal Killing-Yano Tensors}\label{Sec.Definitions}

In this section we shall define and quickly present the main properties of conformal Killing-Yano (CKY) tensors, for more details the reader is refereed to Refs. \cite{Frolov_KY,FrolovBook}. We will work in a Riemannian manifold $(M,\bl{g})$ of dimension $n$, with the signature of the metric $\bl{g}$ being arbitrary. In what follows, $\nabla_a$ denotes the Levi-Civita connection. Square brackets around the tensorial indices means that the enclosed indices are anti-symmetrized, while round brackets denote symmetrization. With this notation, a totally skew-symmetric tensor $\bl{Y}$ of rank $p$ is called a CKY tensor of order $p$ if it obeys the following equation:
\begin{equation}\label{CKY2}
  \nabla_{a}\, Y_{b_1b_2\cdots b_p} \ma \nabla_{b_1}\, Y_{ab_2\cdots b_p} \eq 2\,g_{a[b_1}\,h_{b_2\cdots b_p]} \ma 2\,g_{b_1[a}\,h_{b_2\cdots b_p]} \,,
\end{equation}
where $h_{b_2\cdots b_p}$ is some skew-symmetric tensor of rank $p-1$. It turns out that the above equation is equivalent to the following one:
\begin{equation}\label{CKY1}
  \nabla_{a}\, Y_{b_1b_2\cdots b_p} \eq  \nabla_{[a}\, Y_{b_1b_2\cdots b_p]} \ma 2\,g_{a[b_1}\,h_{b_2\cdots b_p]} \,.
\end{equation}
Contracting the latter equation with $g^{ab_1}$ one can see that $\bl{h}$ is essentially the divergence of $\bl{Y}$,
\begin{equation}\label{h=Divergence}
   h_{b_2\cdots b_p}  \eq \frac{p}{2(n+1-p)}\,\nabla^a\,Y_{ab_2\cdots b_p} \,.
\end{equation}
Although not exploited here, the equation satisfied by a CKY tensor can be nicely cast in terms of differential forms \cite{Kress,Semmelmann,Frolov_KY}. For instance, in \cite{Semmelmann} the latter approach was used to study global aspects of CKY tensors in Sasakian manifolds and to stress the resemblance between these tensors and twistors. Actually, it turns out that twistors can be used to generate CKY tensors \cite{Semmelmann}, while Killing spinors give rise to Killing-Yano tensors \cite{Cariglia}.

The CKY tensors can be used to construct conformal Killing tensors. Indeed, if  $\bl{Y}$ and $\bl{\widetilde{Y}}$ are both conformal Killing-Yano tensors of order $p$ then
\begin{equation}\label{CKT1}
  K_{ab} \eq  Y_{(a}^{\ph{(a}c_2\cdots c_p} \, \widetilde{Y}_{b)c_2\cdots c_p}
\end{equation}
is a conformal Killing tensor of order two. Meaning that the symmetric tensor $\bl{K}$ obeys the following equation:
\begin{equation}\label{CKT-Eq}
  \nabla_{(a}\, K_{bc)} \eq g_{(ab}\,k_{c)}\;,\; \textrm{ with } \; k_c \eq \frac{2}{p} \lef Y_{c}^{\ph{c}d_2\cdots d_p}\,\widetilde{h}_{d_2\cdots d_p} \ma \widetilde{Y}_{c}^{\ph{c}d_2\cdots d_p}\,h_{d_2\cdots d_p} \rig \,.
\end{equation}
In particular, we can take $\bl{Y}=\bl{\widetilde{Y}}$ in (\ref{CKT1}) and (\ref{CKT-Eq}). Therefore, we say that the square of a CKY tensor is a conformal Killing tensor of order two, but the converse generally is not true. The usefulness of conformal Killing tensors relies on the fact that they lead to conserved scalars along null geodesics. Indeed, if $\bl{l}$ is a null affinely parameterized geodesic vector field,
$$ l^a\,l_a \eq 0 \; \textrm{ and }  \; l^a\,\nabla_{a}\, l^b \eq 0\,, $$
then the scalar $C=K_{bc}\,l^bl^c$ is constant along the geodesic curves tangent to $\bl{l}$:
$$   l^a\,\nabla_{a} C \eq l^al^bl^c\,\nabla_a \, K_{bc} \eq l^al^bl^c\,\nabla_{(a} \, K_{bc)} \eq l^al^bl^c\, g_{ab}\,k_c \eq 0\,. $$
Hence, because of (\ref{CKT1}) and (\ref{CKT-Eq}), it follows that CKY tensors are associated with conserved scalars along null geodesics. There are also special types of conformal Killing-Yano tensors that are related to conservation laws along all geodesics, null and non-null, as shown in the remainder of this section. But, before proceeding, it is worth pointing out that besides the totally symmetric conformal Killing tensors and the totally skew-symmetric CKY tensors, higher-rank generalizations of the conformal Killing vectors with indices of non-definite symmetry have also been investigated elsewhere, see \cite{Kress} and references therein.


\subsection{Killing-Yano Tensors}

When the second term on the right hand side of Eq. (\ref{CKY1}) vanishes, $\bl{h}=0$, we say that $\bl{Y}$ is a Killing-Yano (KY) tensor \cite{Yano}. Thus, a KY tensor is just a CKY tensor whose divergence is zero. Note that if $\widetilde{\bl{A}}$ and $\bl{A}$ are both KY tensors of order $p$ then Eqs. (\ref{CKT1}) and (\ref{CKT-Eq}) guarantee that
$$  K_{ab} \eq  A_{(a}^{\ph{(a}c_2\cdots c_p} \, \widetilde{A}_{b)c_2\cdots c_p}  $$
is a Killing tensor, namely $ \nabla_{(a} K_{bc)}=0$. So, if $\bl{t}$ is an affinely parameterized geodesic vector field, $t^a\nabla_at^b=0$, then the scalar $C=K_{ab}t^at^b$ is conserved along the geodesic tangent to $\bl{t}$. Note that $\bl{t}$ need not be null. Moreover, the following skew-symmetric tensor is also conserved along the geodesic:
$$  P_{c_2\cdots c_p} \eq t^b\, A_{bc_2\cdots c_p} \,. $$
Indeed, using (\ref{CKY1}) with $\bl{h}=0$ we find
$$  t^a\,\nabla_a\,P_{c_2\cdots c_p} \eq  t^a t^b \,\nabla_a\, A_{bc_2\cdots c_p} \eq t^a t^b \,\nabla_{[a}\, A_{bc_2\cdots c_p]} \eq 0 \,, $$
where the last equality follows because in the above equation the pair of indices $ab$ is anti-symmetrized while contracted with a symmetric tensor. Note that the conserved scalar $C$ is just the square of this conserved tensor, $C=P^{c_2\cdots c_p}P_{c_2\cdots c_p}$. An extensive account of KY tensors in 4-dimensional spacetimes can be found in Refs. \cite{Dietz,Hall-KY}.


\subsection{Closed Conformal Killing-Yano Tensors}\label{SubSec.CKY1}

We say that $\bl{Y}$ is a closed conformal Killing-Yano tensor whenever the first term on the right hand side of Eq. (\ref{CKY1}) vanishes, $\nabla_{[a} Y_{b_1b_2\cdots b_p]}=0$. These tensors have two very special properties \cite{Frolov_KY}: the Hodge dual of a closed CKY tensor is a KY tensor and, therefore, lead to conservation laws along any geodesic; the exterior product of two closed CKY tensors is another closed CKY tensor. More explicitly, if $\bl{H}$ and $\widetilde{\bl{H}}$ are both closed CKY tensors of order $p$ and $q$ respectively and $\epsilon_{a_1a_2\cdots a_n}$ is the local volume-form of the manifold, then
$$ (H\wedge \widetilde{H})_{a_1a_2\cdots a_p b_1b_2\cdots b_q} \,\equiv\, \frac{(p+q)!}{p!\,q!} \, H_{[a_1a_2\cdots a_p}\,\widetilde{H}_{b_1b_2\cdots b_q]}  $$
is a closed conformal Killing-Yano tensor of order $(p+q)$, and
$$  (\star H)_{b_1b_2\cdots b_{n-p}} \,\equiv\, \frac{1}{p!} \, H_{a_1a_2\cdots a_p}\, \epsilon^{a_1a_2\cdots a_p}_{\ph{a_1a_2\cdots a_p} b_1b_2\cdots b_{n-p}} $$
is a KY tensor of order $(n-p)$. Conversely, one can show that the Hodge dual of a KY tensor is a closed CKY tensor. Since a closed CKY tensor obeys the equation $\nabla_{a}H_{b_1b_2\cdots b_p} = 2g_{a[b_1}\,h_{b_2\cdots b_p]}$, it follows that if $\bl{t}$ is an affinely parameterized geodesic vector field then the tensor
$$ \widehat{P}_{bc_1c_2\cdots c_p} \,\equiv\,  t_{[b}\,H_{c_1c_2\cdots c_p]} $$
is conserved along the orbits of $\bl{t}$. Indeed, on account of $t^a\nabla_at^b=0$, it follows that
$$ t^a\,\nabla_a \,  \widehat{P}_{bc_1c_2\cdots c_p} \eq t^a\,\nabla_a   H_{[c_1c_2\cdots c_p}\,t_{b]} \eq 2\,t^a\, g_{a[c_1}\,h_{c_2\cdots c_p}\,t_{b]} \eq 0 \,.  $$
Particularly, the scalar $\widehat{P}^{bc_1c_2\cdots c_p}\widehat{P}_{bc_1c_2\cdots c_p}$ is also conserved along the geodesic tangent to $\bl{t}$.

Closed CKY tensors proved to be of great relevance in higher-dimensional General Relativity. Indeed, a closed CKY tensor of order two is the central object for achieving integrability of Klein-Gordon, Dirac and geodesic equations in the family of Kerr-NUT-(A)dS spacetimes. In \cite{Frolov_KY,Krtous} it is shown that these spacetimes admit a non-degenerate closed conformal Killing-Yano tensor such that the exterior products of this tensor with itself yield a \textit{tower} of closed CKY tensors. Then, by means of the Hodge dual operation, these tensors are used to construct KY tensors, which, in turn, lead to a set of Killing tensors of order two. The latter objects provide conserved scalars along geodesics, which, eventually, lead to the separability and integrability of the mentioned differential equations.

\section{Integrability Conditions}\label{Sec.IntegrabCond}

In the present section, the integrability conditions for the existence of a conformal Killing-Yano tensor will be worked out. As we shall see, the curvature of the manifold must be constrained in order for Eq. (\ref{CKY2}) to admit a solution. Thus, analysing the integrability conditions one can obtain the possible algebraic types that the curvature must have in order for the space to possess a hidden symmetry. For example, in 4 dimensions a spacetime admits a \emph{non-reducible} Killing-Yano tensor only if the Weyl tensor is either of Petrov type $D$ or vanishes \cite{Steph_KY}. This fact draws our attention to the possibility of type $D$ spacetimes being of relevance. Indeed, spacetimes of type $D$ have proved to be quite special, since Einstein's vacuum equation can be completely integrated in this case \cite{typeD}. Moreover, all known 4-dimensional black-holes are of this type. Since algebraic classifications for the Weyl tensor are now available in any dimension \cite{CMPP,art4,5D class.,Spin6D}, the results of this section paves the way for performing an analogous investigation in manifolds with more than 4 dimensions.

In what follows let us assume that $\bl{Y}$ is a CKY tensor of order $p$, meaning that it obeys Eq. (\ref{CKY2}). Before proceeding to find the integrability conditions, it is helpful to set few definitions. In order to avoid too many indices, we shall define capital indices to be a set of $(p-2)$ indices as follows:
$$  B \,\equiv\, b_3b_4\cdots b_p \quad ; \quad D \,\equiv\, d_3d_4\cdots d_p \,.$$
Using these collective indices, the Ricci identity will be written as:
\begin{equation}\label{RicciId}
  \lef \nabla_a\nabla_b \me  \nabla_b\nabla_a \rig \, Y_{cdE} \eq R_{abc}^{\ph{abc}f}\, Y_{fdE} \ma R_{abd}^{\ph{abd}f}\, Y_{cfE} \ma  R_{ab\hat{e}}^{\ph{abe}\hat{f}} \, Y_{cd\hat{F}}\,,
\end{equation}
where the last term in the above identity amounts to the following expression
\begin{equation}\label{Notation}
  R_{ab\hat{e}}^{\ph{abe}\hat{f}} \,Y_{cd\hat{F}} \,\equiv \, \sum_{i=3}^p \, R_{abe_i}^{\ph{abe_i}f}\,Y_{cde_3\cdots \check{e}_ife_{i+1}\cdots e_p} \,.
\end{equation}
Where, in the latter sum, the notation $\check{e}_i$ means that the index $e_i$ is absent. Now, let us denote the covariant derivative of the tensor $\bl{h}$ defined in Eq. (\ref{CKY2}) by
$$  h'_{a_1a_2B} \eq h'_{a_1a_2b_3\cdots b_p} \,\equiv\, \nabla_{a_1}\, h_{a_2b_3\cdots b_p} \,.$$
Thanks to the skew-symmetry of $\bl{h}$, we have that $\bl{h}'$ obeys $h'_{a_1a_2b_3\cdots b_p}=h'_{a_1[a_2b_3\cdots b_p]}$. Moreover, using this skew-symmetry along with (\ref{h=Divergence}), one can see that $\bl{h}'$ is totally trace-less:
\begin{align*}
h'^a_{\ph{a}\,aC} \,&\propto\, 2\, \nabla^a\nabla^b \, Y_{abC} \eq \lef \nabla^a\nabla^b \me  \nabla^b\nabla^a \rig \, Y_{abC} \\
&\propto\, R^{ab\ph{a}d}_{\ph{ab}a}\, Y_{dbC} \ma R^{ab\ph{b}d}_{\ph{ab}b}\, Y_{adC} \ma \, R^{ab\ph{c}\hat{d}}_{\ph{ab}\hat{c}}\, Y_{ab\hat{D}} \eq 0 \,.
\end{align*}
Where in the latter equality it was used the fact that the Ricci tensor is symmetric as well as the Bianchi identity. Finally, it is also useful to define the following tensor:
\begin{equation}\label{Def.S}
S_{abC} \,\equiv \,  h'_{abC} \ma h'_{baC} \eq \nabla_a\,h_{bC} \ma  \nabla_b\,h_{aC}\,.
\end{equation}
Particularly, note that $\bl{S}$ vanishes if, and only if, $\bl{h}$  is a Killing-Yano tensor of order $(p-1)$. Moreover, since $\bl{h'}$ is trace-less, so is the tensor $\bl{S}$. With these definitions we are ready to move on and find the wanted integrability conditions.

Differentiating Eq. (\ref{CKY2}) and then making a permutation on the indices, we easily find the following relations:
\begin{align*}
   \nabla_a\nabla_b\,Y_{cdE} \ma \nabla_a\nabla_c\,Y_{bdE} &\eq 2\,h'_{a[dE}\,g_{b]c} \ma  2\,h'_{a[dE}\,g_{c]b} \,,  \\
   \nabla_b\nabla_c\,Y_{adE} \ma \nabla_b\nabla_a\,Y_{cdE} &\eq 2\,h'_{b[dE}\,g_{c]a} \ma  2\,h'_{b[dE}\,g_{a]c} \,, \\
   -\nabla_c\nabla_a\,Y_{bdE} \me \nabla_c\nabla_b\,Y_{adE} &\eq -2\,h'_{c[dE}\,g_{a]b} \me  2\,h'_{c[dE}\,g_{b]a}\,.
\end{align*}
Now, summing these three equations and using the Ricci identity (\ref{RicciId}), as well as the Bianchi identity, lead to the following relation:
\begin{align}
 2\,\nabla_a\nabla_b\,Y_{cdE} \eq & 2\, R_{cba}^{\ph{cba}f}\, Y_{fdE} \ma R_{abd}^{\ph{abd}f}\, Y_{cfE} \ma R_{ab\hat{e}}^{\ph{abe}\hat{f}}\, Y_{cd\hat{F}} \ma R_{cad}^{\ph{cad}f}\, Y_{bfE}  \ma R_{ca\hat{e}}^{\ph{cae}\hat{f}}\, Y_{bd\hat{F}} \nonumber \\
 & +\, R_{cbd}^{\ph{cbd}f}\, Y_{afE} \ma R_{cb\hat{e}}^{\ph{cbe}\hat{f}}\, Y_{ad\hat{F}} \ma 2\,h'_{a[dE}\,g_{b]c} \ma 2\,h'_{a[dE}\,g_{c]b} \label{DDY1} \\
 &+\, 2\,h'_{b[dE}\,g_{c]a}  \ma 2\,h'_{b[dE}\,g_{a]c} \me 2\,h'_{c[dE}\,g_{a]b} \me 2\,h'_{c[dE}\,g_{b]a} \,.\nonumber
\end{align}
We can make the above equation more explicit by means of the following algebraic identity
\begin{align*}
  h'_{a[dE}\,g_{b]c} \eq & \frac{1}{p}\, g_{bc} \,h'_{adE} \me \frac{p-1}{p}\, h'_{ab[E}\,g_{d]c} \nonumber\\
  =\, & \frac{1}{p}\, g_{bc} \,h'_{adE} \me \frac{1}{p}\, g_{dc} \,h'_{abE} \ma \, \frac{(p-2)}{p} \,h'_{abd[e_4\cdots e_{p}}\,g_{e_3]c}  \,.
\end{align*}
Using the latter expansion into (\ref{DDY1}) lead us to the following expression for the second derivative of the conformal Killing-Yano tensor $\bl{Y}$:
\begin{align}\label{DDY2}
 2\,\nabla_a\nabla_b\,Y_{cdE} \eq & 2\, R_{cba}^{\ph{bca}f}\, Y_{fdE} \ma R_{abd}^{\ph{abd}f}\, Y_{cfE} \ma R_{ab\hat{e}}^{\ph{abe}\hat{f}}\, Y_{cd\hat{F}} \ma R_{cad}^{\ph{cad}f}\, Y_{bfE}  \ma R_{ca\hat{e}}^{\ph{cae}\hat{f}}\, Y_{bd\hat{F}} \nonumber \\
 & +\, R_{cbd}^{\ph{cbd}f}\, Y_{afE} \ma R_{cb\hat{e}}^{\ph{cbe}\hat{f}}\, Y_{ad\hat{F}} \ma \frac{2}{p} \left[\, 2\,g_{bc}\, h'_{adE} \ma  2\,g_{ac}\, h'_{bdE} \me  2\,g_{ab}\, h'_{cdE} \right. \nonumber \\
 & +\, \left. g_{da} \lef h'_{cbE} - h'_{bcE} \rig \ma g_{db} \lef h'_{caE} - h'_{acE} \rig \me g_{dc} \lef h'_{abE} + h'_{baE} \rig  \, \right]  \\
 & -\,  \frac{2(p-2)}{p}\, \left[ \, h'_{cbd[e_4\cdots e_{p}}\,g_{e_3]a}  \me h'_{bcd[e_4\cdots e_{p}}\,g_{e_3]a} \ma h'_{cad[e_4\cdots e_{p}}\,g_{e_3]b}   \right. \nonumber \\
 &  \ph{2\,\frac{p-2}{p}\,} \quad \left. -\, h'_{acd[e_4\cdots e_{p}}\,g_{e_3]b} \me h'_{abd[e_4\cdots e_{p}}\,g_{e_3]c} \me  h'_{bad[e_4\cdots e_{p}}\,g_{e_3]c} \, \right] \nonumber \,.
\end{align}
Where it is worth recalling that the simplifying notation (\ref{Notation}) was used. Now, just handling (\ref{CKY1}) we can find the following expression:
\begin{align}\label{DD+DD+DD}
 \nabla_a\nabla_b\,Y_{cdE} \ma \nabla_a\nabla_c\,Y_{dbE} \ma \nabla_a\nabla_d\,Y_{bcE} \eq 3\,\nabla_a\nabla_b\,Y_{cdE} \me \frac{6}{p} \, \lef g_{bc} \, h'_{adE} \me   g_{bd} \, h'_{acE}\rig \nonumber\\
  +\, \frac{2(p-2)}{p} \lef h'_{adb[e_4\cdots e_{p}}\,g_{e_3]c} \ma h'_{abc[e_4\cdots e_{p}}\,g_{e_3]d} \me 2\, h'_{acd[e_4\cdots e_{p}}\,g_{e_3]b} \rig\,.
\end{align}
Then, taking advantage of Eq. (\ref{DDY2}), and its copies with the indices $bcd$ permuted, to rewrite the left hand side of (\ref{DD+DD+DD}), we end up with
\begin{align}\label{DDY3}
 2\,\nabla_a\nabla_b\, Y_{cdE}  &\eq  -3\, R^f_{\ph{f}a[bc}\, Y_{d]fE} \me  R^{\hat{f}}_{\ph{f}\hat{e}[bc}\, Y_{d]a\hat{F}} \me 2\,
 R^{\hat{f}}_{\ph{f}\hat{e}a[b}\, Y_{cd]\hat{F}} \nonumber \\
 & +\,  \frac{4}{p} \lef g_{bc}\, h'_{adE} \me  g_{bd}\, h'_{acE} \me 3\,  g_{a[b}\, h'_{cd]E}   \rig   \\
 & -\, \frac{2(p-2)}{p} \lef -2\, h'_{[bcd][e_4\cdots e_{p}}\,g_{e_3]a}  \ma h'_{[abd][e_4\cdots e_{p}}\,g_{e_3]c} \ma h'_{[acb][e_4\cdots e_{p}}\,g_{e_3]d}  \right. \nonumber \\
 & +\, \left. h'_{[adc][e_4\cdots e_{p}}\,g_{e_3]b} \ma  h'_{adb[e_4\cdots e_{p}}\,g_{e_3]c} \ma  h'_{abc[e_4\cdots e_{p}}\,g_{e_3]d} \ma  h'_{adc[e_4\cdots e_{p}}\,g_{e_3]b} \rig \nonumber \,.
\end{align}
Now, equating the right hand sides of Eqs. (\ref{DDY2}) and (\ref{DDY3}) we find, after some algebra, the following relation:
\begin{align}\label{IntCond1}
 0  \eq &   R^f_{\ph{f}acb}\, Y_{dfE} \ma R^f_{\ph{f}bda}\, Y_{cfE} \ma R^f_{\ph{f}cad}\, Y_{bfE} \ma R^f_{\ph{f}dbc}\, Y_{afE} \ma
  R^{\hat{f}}_{\ph{f}\hat{e}[bc}\, Y_{d]a\hat{F}}     \ma 2\, R^{\hat{f}}_{\ph{f}\hat{e}a[b}\, Y_{cd]\hat{F}}  \nonumber\\
 & +\, 2\, R^{\hat{f}}_{\ph{f}\hat{e}bc}\, Y_{ad\hat{F}} \me 3\, R^{\hat{f}}_{\ph{f}\hat{e}[ab}\, Y_{c]d\hat{F}} \ma \frac{2}{p} \lef g_{ac}\, S_{bdE} \ma
  g_{bd}\, S_{acE} \me g_{ab}\, S_{cdE} \me g_{cd}\, S_{abE} \right) \nonumber\\
  & +\, \frac{2(p-2)}{3\,p}\, \left(\, S_{dcb[e_4\cdots e_{p}}\,g_{e_3]a} \me S_{dbc[e_4\cdots e_{p}}\,g_{e_3]a} \ma S_{cda[e_4\cdots e_{p}}\,g_{e_3]b} \me S_{cad[e_4\cdots e_{p}}\,g_{e_3]b} \right. \nonumber\\
  & +\, \left. S_{bad[e_4\cdots e_{p}}\,g_{e_3]c} \me S_{bda[e_4\cdots e_{p}}\,g_{e_3]c} \ma S_{abc[e_4\cdots e_{p}}\,g_{e_3]d} \me
  S_{acb[e_4\cdots e_{p}}\,g_{e_3]d}  \,\, \right) \,.
\end{align}
Where the tensor $\bl{S}$ used above was defined in (\ref{Def.S}). Equation (\ref{IntCond1}) is not the integrability condition yet, since $\bl{S}$ is defined in terms of the second derivative of $\bl{Y}$. However, contracting (\ref{IntCond1}) with $g^{ab}$, one can find an expression for $\bl{S}$ depending just on the Riemann tensor and $\bl{Y}$, without derivatives. The final result is:
\begin{equation}\label{S=RicciY+RiemannY}
  S_{cdE} \eq \frac{p}{n-p} \left[\, R^f_{\ph{f}(c}\,Y_{d)fE} \ma \frac{1}{2}\,R^{a\hat{f}}_{\ph{af}\hat{e}(c}\,Y_{d)a\hat{F}} \, \right] \,,
\end{equation}
with $R^a_{\ph{a}b}\equiv R^{ca}_{\ph{ca}cb}$ standing for the Ricci tensor. Inserting this relation into (\ref{IntCond1}) we finally arrive at the integrability condition for $\bl{Y}$ to be a conformal Killing-Yano tensor of order $p$. Note that such integrability condition amounts to an algebraic constraint for the Riemann tensor. It is worth stressing that albeit this integrability condition is necessary for the existence of a CKY tensor, it is not sufficient. Although the right hand side of (\ref{S=RicciY+RiemannY}) diverges when $n=p$, it is not necessary to worry about this case since every non-zero $n$-form is a CKY tensor. Therefore, the existence of a CKY tensor of order $p=n$ represents no local constraint. Expanding the Riemann tensor in terms of the Weyl tensor, the Ricci tensor and the Ricci scalar, we can put Eq. (\ref{S=RicciY+RiemannY}) in the following form:
\begin{equation}\label{S=RicciY+WeylY}
  S_{cdE} \eq \frac{p}{n-2} \, R^f_{\ph{f}(c}\,Y_{d)fE} \ma \frac{p}{2(n-p)} \,C^{a\hat{f}}_{\ph{af}\hat{e}(c}\,Y_{d)a\hat{F}}  \,,
\end{equation}
with $C^{ab}_{\ph{ab}cd}$ standing for the Weyl tensor. Note that the case $p=1$  is not encompassed by the calculations performed in this section. However, in such a case $\bl{Y}$ is a conformal Killing vector and does not represent a \emph{hidden} symmetry. Anyhow, the integrability conditions for the existence of a closed conformal Killing vector can be found in \cite{Bat-KYn-1}. Thus, in what follows let us assume $p\geq 2$.


\subsection{Invariance Under Conformal Transformations}\label{Sub.Sec-ConformalTransf}

It turns out that the equation satisfied by a conformal Killing-Yano tensor, Eq. (\ref{CKY2}), is invariant under conformal transformations. More precisely, one can prove that if $Y_{a_1\cdots a_p}$ is a CKY tensor of order $p$ in the manifold $(M,g_{ab})$ then the tensor $\Omega^{p+1}Y_{a_1\cdots a_p}$ is a CKY tensor in the manifold $(M,\Omega^2g_{ab})$, where $\Omega$ is any non-vanishing function. Therefore, we should expect the integrability condition for the existence of a conformal Killing-Yano tensor to be invariant under conformal transformations. Thus, since the conformally invariant part of the Riemann tensor is the Weyl tensor, it is reasonable that such integrability condition could be expressed just in terms of the Weyl tensor and the CKY tensor. Indeed, expanding the Riemann tensor in Eq. (\ref{IntCond1}) in terms of the Weyl tensor, the Ricci tensor, and the Ricci scalar and inserting (\ref{S=RicciY+WeylY}) into (\ref{IntCond1}), one can prove that the terms containing the Ricci tensor and the Ricci scalar are canceled out, so that we are left with an integrability condition involving just the Weyl tensor. More explicitly, the integrability condition (\ref{IntCond1}) is equivalent to the following analogous equation:
\begin{align}\label{IntCond-Weyl}
 0  \eq &   C^f_{\ph{f}acb}\, Y_{dfE} \ma C^f_{\ph{f}bda}\, Y_{cfE} \ma C^f_{\ph{f}cad}\, Y_{bfE} \ma C^f_{\ph{f}dbc}\, Y_{afE} \ma
  C^{\hat{f}}_{\ph{f}\hat{e}[bc}\, Y_{d]a\hat{F}}     \ma 2\, C^{\hat{f}}_{\ph{f}\hat{e}a[b}\, Y_{cd]\hat{F}}  \nonumber\\
 & +\, 2\, C^{\hat{f}}_{\ph{f}\hat{e}bc}\, Y_{ad\hat{F}} \me 3\, C^{\hat{f}}_{\ph{f}\hat{e}[ab}\, Y_{c]d\hat{F}} \ma \frac{2}{p} \lef g_{ac}\, W_{bdE} \ma
  g_{bd}\, W_{acE} \me g_{ab}\, W_{cdE} \me g_{cd}\, W_{abE} \right) \nonumber \\
  & +\, \frac{2(p-2)}{3\,p}\, \left[\, W_{dcb[e_4\cdots e_{p}}\,g_{e_3]a} \me W_{dbc[e_4\cdots e_{p}}\,g_{e_3]a} \ma W_{cda[e_4\cdots e_{p}}\,g_{e_3]b} \me W_{cad[e_4\cdots e_{p}}\,g_{e_3]b} \right. \nonumber\\
  & +\, \left. W_{bad[e_4\cdots e_{p}}\,g_{e_3]c} \me W_{bda[e_4\cdots e_{p}}\,g_{e_3]c} \ma W_{abc[e_4\cdots e_{p}}\,g_{e_3]d} \me
  W_{acb[e_4\cdots e_{p}}\,g_{e_3]d}  \, \right] ,
\end{align}
where $W_{cdE}$ is the part of $S_{cdE}$ in Eq.(\ref{S=RicciY+WeylY}) that is conformally invariant,
\begin{equation*}\label{W-def}
  W_{cdE} \,\equiv\,   \frac{p}{2(n-p)} \,C^{a\hat{f}}_{\ph{af}\hat{e}(c}\,Y_{d)a\hat{F}} \,.
\end{equation*}
Now, let us explore some specific cases of the above development.


\subsection{The Case $p=2$}\label{Sub.Sec.p2}

Suppose now that $p=2$. Since in this circumstance $\bl{Y}$ has just two indices, in the above expressions we shall ignore the terms such that the indices $e_i$ appear in the Riemann tensor or in the metric. So, just the free indices $a$, $b$, $c$ and $d$ should be present when $p=2$. Taking this into account, the Eqs. (\ref{DDY3}), (\ref{IntCond1}) and (\ref{S=RicciY+RiemannY}) become respectively:
$$  2\,\nabla_a\nabla_b\, Y_{cd}  \eq  -3\, R^f_{\ph{f}a[bc}\, Y_{d]f} \ma 2 \lef g_{bc}\, h'_{ad} \me  g_{bd}\, h'_{ac} \me 3\,  g_{a[b}\, h'_{cd]} \rig \,, $$
\begin{align}\label{p2-IC}
 0  \eq &    R^f_{\ph{f}acb}\, Y_{df} \ma R^f_{\ph{f}bda}\, Y_{cf} \ma R^f_{\ph{f}cad}\, Y_{bf} \ma R^f_{\ph{f}dbc}\, Y_{af} \nonumber\\
 & \ma \lef g_{ac}\, S_{bd} \ma
  g_{bd}\, S_{ac} \me g_{ab}\, S_{cd} \me g_{cd}\, S_{ab} \right) \,,
\end{align}
\begin{align}\label{S-p2}
   S_{cd} \eq \frac{2}{n-2} \, R^f_{\ph{f}(c}\,Y_{d)f} \,.
\end{align}
These equations are in perfect agreement with the ones proved by Tachibana in \cite{Tachibana-CKY}. As explained in Sec. \ref{Sub.Sec-ConformalTransf}, inserting (\ref{S-p2}) into (\ref{p2-IC}) we find that the Ricci tensor and the Ricci scalar are canceled out, so that we end up with the following integrability condition:
\begin{equation}\label{Int.Cond.Weyl-p2}
  0  \eq     C^f_{\ph{f}acb}\, Y_{df} \ma C^f_{\ph{f}bda}\, Y_{cf} \ma C^f_{\ph{f}cad}\, Y_{bf} \ma C^f_{\ph{f}dbc}\, Y_{af}
\end{equation}

Now, let us investigate which restrictions the integrability condition (\ref{Int.Cond.Weyl-p2}) imposes over the algebraic type of the Weyl tensor in 4-dimensional manifolds of arbitrary signature. For this purpose it is useful to use spinorial language \cite{PenroseBook,Plebanski75}. Note that Eq. (\ref{Int.Cond.Weyl-p2}) can be written as
$$ G_{abcd} \eq 0\;,\; \textrm{ where }\; G_{abcd}\, \equiv \, C_{ab[c}^{\ph{ab[c}f}\,Y_{d]f} \ma  C_{cd[a}^{\ph{cd[a}f}\,Y_{b]f} \,.$$
The interesting thing about the tensor $\bl{G}$ is that it has the same algebraic symmetries of the Weyl tensor:
$$  G_{abcd} \eq G_{[ab][cd]} \;\,, \quad G_{abcd} \eq G_{cdab} \;\,, \quad G_{a[bcd]} \eq 0 \quad  \textrm{and}\quad  G^a_{\ph{a}bad} \eq 0 \,.  $$
Therefore, just as the spinorial representation of the self-dual part of the Weyl tensor in 4-dimensional manifolds is $\Psi_{\alpha\beta\rho\sigma}=\Psi_{(\alpha\beta\rho\sigma)}$, the spinorial representation of the self-dual part of $\bl{G}$ must, likewise, have four totally symmetric indices, $\Gamma_{\alpha\beta\rho\sigma}=\Gamma_{(\alpha\beta\rho\sigma)}$. With the Greek indices being spinorial indices ranging from one to two. Since $\bl{G}$ is a contraction of the Weyl tensor with $\bl{Y}$, by lack of other possibilities, we must have
$$ \Gamma_{\alpha\beta\rho\sigma} \,\propto\, \Psi_{\kappa(\alpha\beta\rho} \, \Upsilon_{\sigma)}^{\ph{\sigma)}\kappa} \,. $$
With $\Upsilon_{\alpha\beta}=\Upsilon_{(\alpha\beta)}$ denoting the spinorial equivalent of the self-dual part of the bivector $Y_{ab}$. Thus, the integrability condition for the existence of a CKY tensor is
\begin{equation}\label{Int.Cond.Spinor-p2}
 \Psi_{\kappa(\alpha\beta\rho} \, \Upsilon_{\sigma)}^{\ph{\sigma)}\kappa}  \eq 0
\end{equation}
along with the anti-self-dual analog of this equation. In a 4-dimensional manifold, the self-dual part of a bivector can have just two algebraic types, it can either be null or non-null. In the former case its spinorial representation in a suitable frame is $\Upsilon_{\alpha\beta}= o_\alpha o_\beta$. Inserting this into (\ref{Int.Cond.Spinor-p2}) we are led to the conclusion that $o^\alpha\,\Psi_{\alpha\beta\rho\sigma}=0$, which implies that either the self-dual part of the Weyl tensor vanishes or its Petrov type is $N$. On the other hand, if the self-dual part of the bivector is non-null we have that $\Upsilon_{\alpha\beta}\propto o_{(\alpha} \iota_{\beta)}$, where $\iota^\alpha o_\alpha=1$. Inserting this expression for $\Upsilon_{\alpha\beta}$ into (\ref{Int.Cond.Spinor-p2}) we find that either $\Psi_{\alpha\beta\rho\sigma}$ vanishes or its only non-zero component is $o^\alpha o^\beta\,\Psi_{\alpha\beta\rho\sigma}\, \iota^\rho \iota^\sigma$. In the latter case the self-dual part of the Weyl tensor is of Petrov type $D$ . Since an analogous analysis holds for the anti-self-dual part of the Weyl tensor, we conclude that if a 4-dimensional manifold admits a CKY tensor of order two then the algebraic type of the Weyl tensor, in the notation of Ref. \cite{art1}, must be one of the following: $(O,O)$, $(O,N)$, $(O,D)$, $(N,N)$, $(N,D)$ or $(D,D)$. Particularly, using the results of \cite{art1}, this implies that if a 4-dimensional Lorentzian spacetime admits a CKY tensor of order two then the Petrov type of the Weyl tensor is $O$, $N$ or $D$, a fact that has already been established before \cite{Glass}. This also generalizes the results obtained in \cite{Collinson,Steph_KY} for KY tensors in Lorentzian spacetimes. In the same vein, one conclude that if a 4-dimensional manifold of Euclidean signature admits a CKY tensor of order two then the algebraic type of the Weyl tensor might be $(O,O)$, $(O,D)$ or $(D,D)$. For a review about the Petrov classification see \cite{Bat-Book}.



\subsection{The Case $p=3$}

Since the cases $p=1$ and $p=2$ have been widely considered before in the literature \cite{Bat-KYn-1,Tachibana-CKY}, the case $p=3$ is the first relevant case in this article. Therefore, it is worth making the previous equations explicit in this particular case. Assuming $p=3$ in Eqs. (\ref{DDY3}), (\ref{IntCond1}) and (\ref{S=RicciY+RiemannY}), we find the following equations respectively.
\begin{align*}
 2\,\nabla_a\nabla_b\, Y_{cde}   \eq & -3\, R^f_{\ph{f}a[bc}\, Y_{d]fe} \me  R^{f}_{\ph{f}e[bc}\, Y_{d]af} \me 2\,
 R^{f}_{\ph{f}ea[b}\, Y_{cd]f} \\
 & +\,  \frac{2}{3} \,\left[\, 2\,g_{bc}\, h'_{ade} \me   2\,g_{bd}\, h'_{ace} \me  6\,  g_{a[b}\, h'_{cd]e} \ma (\, h'_{abd}  \me  h'_{[abd]} \,)\, g_{ec}  \right.   \\
 & + \left. \,(\, h'_{acb} \me h'_{[acb]} \,)\,g_{ed}  \ma (\,  h'_{acd} \ma  h'_{[acd]} \,)\, g_{eb}  \ma 2\, h'_{[bcd]}\,g_{ea} \,\right]  \,,
\end{align*}
\begin{align}
 0  &\eq    C^f_{\ph{f}acb}\, Y_{dfe} \ma C^f_{\ph{f}bda}\, Y_{cfe} \ma C^f_{\ph{f}cad}\, Y_{bfe} \ma C^f_{\ph{f}dbc}\, Y_{afe} \ma
  C^{f}_{\ph{f}e[bc}\, Y_{d]af}     \ma 2\, C^{f}_{\ph{f}ea[b}\, Y_{cd]f}  \nonumber\\
 & +\, 2\, C^{f}_{\ph{f}ebc}\, Y_{adf} \me 3\, C^{f}_{\ph{f}e[ab}\, Y_{c]df} \ma \frac{2}{3} \lef g_{ac}\, W_{bde} \ma
  g_{bd}\, W_{ace} \me g_{ab}\, W_{cde} \me g_{cd}\, W_{abe} \right)  \label{IntCondp3}\\
  & +\, \frac{2}{9}\, \left[\, \lef W_{dcb} \me W_{dbc}\rig g_{ea} \ma \lef W_{cda} \me W_{cad} \rig g_{eb}
  \ma \lef W_{bad} \me W_{bda}\rig g_{ec} \ma \lef W_{abc} \me   W_{acb}\rig g_{ed}  \, \right] \,, \nonumber
\end{align}
\begin{align}\label{W-p3}
W_{cde} \eq  \frac{3}{2(n-3)} \,C^{af}_{\ph{af}e(c}\,Y_{d)af}  \,.
\end{align}


\subsubsection{Comparison with Kashiwada's result}\label{Sub.Sub.Sec.Kashiwada}

As mentioned in the introductory section, an integrability condition for CKY tensors of arbitrary rank have also been worked out by Kashiwada in \cite{Kashiwada-Int}. In what follows, the latter integrability condition will be compared with the one obtained here for the case $p=3$. If $\bl{Y}$ is a rank 3 CKY tensor then Kashiwada's result states that the following condition must hold:
$$   R^{f}_{\ph{f}b[da}\,Y_{e]fc}  \me R^{f}_{\ph{f}c[da}\,Y_{e]fb}\me  2\,R_{cb\ph{f}[d}^{\ph{cb}f}\,Y_{ae]f}  \ma \frac{4}{3}\lef S_{b[da}\,g_{e]c} \me S_{c[da}\,g_{e]b} \rig \eq 0  \,,$$
where
$$ S_{cde} \eq \frac{3}{n-2} \, R^f_{\ph{f}(c}\,Y_{d)fe} \ma \frac{3}{2(n-3)} \,C^{af}_{\ph{af}e(c}\,Y_{d)af}  \,. $$
However, as argued in Section \ref{Sub.Sec-ConformalTransf}, since the CKY equation is conformally invariant it follows that its integrability condition can be expressed just in terms of the Weyl tensor, a fact that was not pointed out in \cite{Kashiwada-Int}. In particular, one can check that Kashiwada's integrability condition can be written as:
\begin{equation}\label{Kashiwada-IC}
   C^{f}_{\ph{f}b[da}\,Y_{e]fc}  \me C^{f}_{\ph{f}c[da}\,Y_{e]fb} \me  2\,C_{cb\ph{f}[d}^{\ph{cb}f}\,Y_{ae]f}\ma \frac{4}{3}\lef W_{b[da}\,g_{e]c} \me W_{c[da}\,g_{e]b} \rig \eq 0\,,
\end{equation}
with $\bl{W}$ given by (\ref{W-p3}). Anti-symmetrizing the indices $ade$ in Eq. (\ref{IntCondp3})  one verify that Eq. (\ref{Kashiwada-IC}) is readily obtained. This proves that Kashiwada's result is contained in the integrability condition obtained here. Conversely, adding (\ref{Kashiwada-IC}) to its permutation obtained by means of the change $\{a,b\}\leftrightarrow\{c,d\}$, one arrive at (\ref{IntCondp3})  \footnote{Throughout these manipulations it is useful to make use of some algebraic properties of the tensor $W_{abc}$. Besides the identity $W_{abc}=W_{(ab)c}$, it follows from (\ref{W-p3}) that $W_{(abc)}=0$. As a consequence, one can prove that $W_{a(bc)}\eq -\frac{1}{2} W_{bca}$ and $ W_{a[bc]}\ma W_{b[ac]} \eq \frac{3}{2} W_{abc}$.}. Therefore, Kashiwada's integrability condition is equivalent to the one obtained here. It is interesting noting that whereas here just four indices of the second derivative of the CKY tensor are permuted in order to obtain the integrability condition, irrespective of the order of the CKY tensor, in Kashiwada's deduction all free indices are treated on the same footing by means of antisymmetric permutations.  Therefore, it would be natural for the integrability condition obtained here to be more general than the one obtained by Kashiwada. However, as we have just seen, both integrability conditions turn out to be equivalent. It is also worth noting that in Ref. \cite{Semmelmann}  Semmelmann have recast Kashiwada's integrability condition using differential forms, see also \cite{Kora}.


\subsection{Killing-Yano Tensors}\label{Sub.Sec-KY2}

A Killing-Yano tensor is a conformal Killing-Yano tensor whose divergence vanishes. So, according to (\ref{h=Divergence}), $\bl{Y}$ is a KY tensor if, and only if, $\bl{h}$ vanishes identically. Thus, in order to obtain the integrability condition for a KY tensor of order $p$ we just need to plug $\bl{h}=0$ into the previous equations. In particular, this means that $\bl{h}'$ and $\bl{S}$ are both zero. Hence, Eqs. (\ref{IntCond1}) and (\ref{S=RicciY+WeylY}) yield the following integrability conditions for a KY tensor $\bl{A}$ of order $p\geq2$:
\begin{align*}
 0   \,= & \, \,   R^f_{\ph{f}acb}\, A_{dfE} \ma R^f_{\ph{f}bda}\, A_{cfE} \ma R^f_{\ph{f}cad}\, A_{bfE} \ma R^f_{\ph{f}dbc}\, A_{afE} \nonumber \\
  & +\, R^{\hat{f}}_{\ph{f}\hat{e}[bc}\, A_{d]a\hat{F}}     \ma 2\, R^{\hat{f}}_{\ph{f}\hat{e}a[b}\, A_{cd]\hat{F}}  \ma
 2\, R^{\hat{f}}_{\ph{f}\hat{e}bc}\, A_{ad\hat{F}} \me 3\, R^{\hat{f}}_{\ph{f}\hat{e}[ab}\, A_{c]d\hat{F}} \,\;,
\end{align*}
\begin{align*}
0 \eq   2\,(n-p)\, R^f_{\ph{f}(c}\,A_{d)fE} \ma (n-2)\,  C^{a\hat{f}}_{\ph{af}\hat{e}(c}\,A_{d)a\hat{F}} \,.
\end{align*}
In addition, Eq. (\ref{DDY3}) implies that the second derivative of a Killing-Yano tensor is
\begin{align*}
 2\,\nabla_a\nabla_b\, A_{cdE}  &\eq  -3\, R^f_{\ph{f}a[bc}\, A_{d]fE} \me  R^{\hat{f}}_{\ph{f}\hat{e}[bc}\, A_{d]a\hat{F}} \me 2\,
 R^{\hat{f}}_{\ph{f}\hat{e}a[b}\, A_{cd]\hat{F}} \,.
\end{align*}
Since $\bl{A}$ is a KY tensor, it follows that the left hand side of the above equation is totally anti-symmetric in the indices $bcdE$. Therefore, we have the right to skew-symmetrize these indices on the right hand side of such equation. Performing this anti-symmetrization and using the Bianchi identity, we eventually arrive at the following expression for the second derivative of a KY tensor:
\begin{align}\label{DDY-KY}
 2\,\nabla_a\nabla_b\, A_{cdE}  &\eq  -\,(p+1) \,  A_{f[dE} \, R_{bc]a}^{\ph{bc]a}f} \,,
\end{align}
a relation that has already been obtained before by Tachibana and Kashiwada in \cite{Tachibana-KY2}, see also \cite{Cariglia}.


\subsection{Closed Conformal Killing-Yano Tensors}

As reasoned in Sec. \ref{SubSec.CKY1}, another special class of conformal Killing-Yano tensors is formed by the closed CKY tensors. These are skew-symmetric tensors $\bl{H}$ that obey the following equation:
\begin{equation}\label{CCKY}
    \nabla_{a}\, H_{b_1b_2\cdots b_p} \eq   2\,g_{a[b_1}\,h_{b_2\cdots b_p]} \,.
\end{equation}
Then, taking the covariant derivative of this equation we find:
$$   \nabla_{a}\nabla_{b}\, H_{cdE} \eq   2\,h'_{a[dE}\,g_{c]b} \,. $$
Using the above equation along with the Ricci identity (\ref{RicciId}) lead to the following relation:
\begin{equation}\label{Int.Cond-CCKY}
 R_{abc}^{\ph{abc}f}\, H_{fdE} \ma R_{abd}^{\ph{abd}f}\, H_{cfE} \ma  R_{ab\hat{e}}^{\ph{abe}\hat{f}} \, H_{cd\hat{F}} \me  2\,h'_{a[dE}\,g_{c]b} \ma 2\,h'_{b[dE}\,g_{c]a} \eq 0 \,.
\end{equation}
Finally, contracting (\ref{Int.Cond-CCKY}) with $g^{ac}$ yield
\begin{equation}\label{h'CCKY}
 h'_{bdE} \eq \frac{p}{4(n-p)} \lef 2\, R_b^{\ph{b}f}\,H_{dfE} \ma R^{af}_{\ph{af}bd}\,H_{afE} \ma \,R^{a\hat{f}}_{\ph{af}b\hat{e}}\,H_{ad\hat{F}}  \rig \,.
\end{equation}
Inserting (\ref{h'CCKY}) into (\ref{Int.Cond-CCKY}) lead to the integrability condition for a closed CKY tensor of order $p$. So, compared with a general CKY tensor, the integrability condition is much simpler in the closed case. Taking the symmetric part of the pair of indices $bd$ in (\ref{h'CCKY}), we easily see that the expression for $\bl{S}$ is the same as in the general case, see (\ref{Def.S}) and (\ref{S=RicciY+RiemannY}). Another (equivalent) way of analysing the integrability conditions for the existence of a closed CKY tensor is to use the fact that its Hodge dual is a KY tensor an then employ the results of Sec. \ref{Sub.Sec-KY2}. It is worth pointing out that in \cite{Houri-Uniquiness} it has been proven that a spacetime admitting a non-degenerate CKY tensor of order two must be contained in the Kerr-NUT-(A)dS class.

\section{Constructing Killing-Yano Tensors in Einstein Spaces}\label{Sec.EinsteinSpaces}

An Einstein space is a manifold such that the Ricci tensor is proportional to the metric, $R_{ab}=\Lambda g_{ab}$. Because of the contracted Bianchi identity, it follows that $\Lambda$ is necessarily constant. Physically, these spaces represent solutions of Einstein's vacuum equation with a cosmological constant. It turns out that, in an Einstein space, given a CKY tensor of order $p=2$ one can construct a Killing vector. Indeed, using (\ref{S-p2}) we see that if $\bl{Y}$ is a CKY tensor of order two in an Einstein space then
$$  S_{cd} \eq \frac{1}{n-2} \, \lef R^f_{\ph{f}c}\,Y_{df} \ma R^f_{\ph{f}d}\,Y_{cf} \rig \eq  \frac{\Lambda}{n-2}\lef  Y_{dc} \ma Y_{cd}\rig \eq 0 \,. $$
Since $\bl{S}$ was defined in Eq. (\ref{Def.S}) by $S_{ab}= \nabla_{a}h_b + \nabla_{b}h_a$ when $p=2$, it follows from the above equation that $h_a$ is a Killing vector field. Where it was assumed that $h_a\neq0$, namely $\bl{Y}$ is not a KY tensor. This fact was first proved in \cite{Tachibana-CKY} and it was of fundamental importance for the construction of the \textit{tower} of symmetries in the Kerr-NUT-(A)dS spacetimes of arbitrary dimension \cite{Frolov_KY}.

Since a Killing vector is a KY tensor of order $p=1$, in the preceding paragraph we showed that, in an Einstein space, the divergence of a CKY tensor of order two is a KY tensor of order one. So, a natural question to be posed is the following: If $Y_{a_1a_2\cdots a_p}$ is a CKY tensor of order $p>2$ in an Einstein space then is $h_{a_2\cdots a_p}$, defined in (\ref{h=Divergence}), a KY tensor of order $(p-1)$? The answer is generally no, since Eqs. (\ref{Def.S}) and (\ref{S=RicciY+WeylY}) imply that in an Einstein space we have
$$   \nabla_ch_{dE} \ma \nabla_dh_{cE} \eq \frac{p}{2(n-p)} \,C^{a\hat{f}}_{\ph{af}\hat{e}(c}\,Y_{d)a\hat{F}}  \,.  $$
Therefore, $\bl{h}$ is a Killing-Yano tensor if, and only if,
\begin{equation}\label{KY-Cond-EinsteinS}
  C^{a\hat{f}}_{\ph{af}\hat{e}(c}\,Y_{d)a\hat{F}} \eq 0 \,,
\end{equation}
where we shall assume that $\bl{Y}$ is not a KY tensor, namely $\bl{h}\neq 0$. Equation (\ref{KY-Cond-EinsteinS}) represents an algebraic constraint for the  Weyl tensor. Therefore, differently from the case $p=2$, when $p>2$ the divergence of a CKY tensor in an Einstein space is a KY tensor only in manifolds with algebraically constrained Weyl tensors. In principle, one could reason that condition (\ref{KY-Cond-EinsteinS}) could be just a consequence of the integrability condition (\ref{IntCond-Weyl}). However, after working with Eq. (\ref{IntCond-Weyl}), it seems that the latter possibility does not hold. Thus, it would be valuable to investigate what Eq. (\ref{KY-Cond-EinsteinS}) implies in terms of the algebraic classifications of Refs. \cite{CMPP,art4,5D class.,Spin6D}. As an example, let us work out these implications for the case $p=3$ in 4-dimensional spacetimes.

\subsection{The Case $p=3$ in 4 Dimensions}

Let $Y_{abc}$ be a CKY tensor with non-vanishing divergence in an Einstein space. Then, according to (\ref{KY-Cond-EinsteinS}), the skew-symmetric tensor
$$ h_{ab} \eq  \frac{3}{2(n-2)}\,\nabla^a\,Y_{abc} $$
is a KY tensor if, and only if,
\begin{equation*}\label{Einstein-p3-1}
  C^{de}_{\ph{de}a(b}\,Y_{c)de} \eq 0 \,.
\end{equation*}
It is simple matter to prove that the latter condition is tantamount to
\begin{equation}\label{Einstein-p3-2}
 C^{de}_{\ph{de}ab}\,Y_{cde} \eq  C^{de}_{\ph{de}[ab}\,Y_{c]de} \,.
\end{equation}
Interestingly, the term on the right hand side of (\ref{Einstein-p3-2}) is just the action of the \emph{Weyl operator} on the 3-form $Y_{abc}$, see Ref. \cite{art4}. Therefore, it is reasonable to expect that the algebraic classification defined in \cite{art4} plays an important role in the analysis of condition (\ref{Einstein-p3-2}). For instance, it was proved in \cite{art4} that in 4 dimensions the action of the \emph{Weyl operator} in 3-forms gives zero. So, in the particular case $n=4$, Eq. (\ref{Einstein-p3-2}) implies that
\begin{equation}\label{Einstein-p3-4D1}
  C^{de}_{\ph{de}ab}\,Y_{cde} \eq 0 \,.
\end{equation}
Instead of analysing condition (\ref{Einstein-p3-4D1}) in terms of $\bl{Y}$, it is fruitful to use its Hodge dual. Defining the 1-form $\bl{\xi}$ by $Y_{bcd}=\xi^a\epsilon_{abcd}$, it follows that (\ref{Einstein-p3-4D1}) is equivalent to:
\begin{equation}\label{Einstein-p3-4D2}
  C_{ab}^{\ph{ab}[cd}\,\xi^{e]} \eq 0 \,.
\end{equation}
In particular, contracting the indices $a$ and $e$ in (\ref{Einstein-p3-4D2}), we find that $\xi^aC_{abcd}=0$. Conversely, it turns out that in Lorentzian 4-dimensional manifolds the latter condition implies (\ref{Einstein-p3-4D2}). Furthermore, if $\xi^aC_{abcd}=0$ then either the Weyl tensor vanishes or its Petrov type is $N$ with $\bl{\xi}$ being the repeated principal null direction \cite{Bat-Book}. Therefore, we have just proved that if $\bl{Y}$ is a CKY tensor of order three in a 4-dimensional Einstein manifold of Lorentzian signature and non-vanishing Weyl tensor, then its divergence is a KY tensor of order two if, and only if, the Weyl tensor is type $N$ with the Hodge dual of $\bl{Y}$ being the repeated principal null direction.

\section{Maximally Symmetric Spaces}\label{Sec.MaximallySym}

In this section we shall prove some interesting results concerning CKY tensors in maximally symmetric spaces. These results are generalizations of the ones obtained in \cite{Tachibana-CKY} for the particular case $p=2$. This attempt of generalizing the results of \cite{Tachibana-CKY} to arbitrary $p$ have also been pursued by Kashiwada in \cite{Kashiwada-Int}. A manifold of dimension $n$ is called a maximally symmetric space when it admits the maximum number of isometries, which means that it has $\frac{1}{2}n(n+1)$ independent Killing vector fields. For instance, the de Sitter and Anti-de Sitter spacetimes are maximally symmetric manifolds of Lorentzian signature. The Riemann tensor of a maximally symmetric space is given by
\begin{equation}\label{Riemann-MaxSym}
 R_{abcd} \eq \lambda \, \lef g_{ac}\,g_{bd} \me  g_{ad}\,g_{bc} \rig \,,
\end{equation}
where $\lambda$ is some constant scalar. Equivalently, a maximally symmetric manifold is characterized as being an Einstein space that is conformally flat,
\begin{equation}\label{MaxSym-RicciWeyl}
  R_{ab} \eq (n-1)\,\lambda\, g_{ab} \quad  \textrm{ and } \quad C_{abcd} \eq 0 \,.
\end{equation}
Since the case $\lambda=0$ represents the trivial flat space, in what follows it is assumed  $\lambda\neq 0$.

Now, let $\bl{\mathcal{A}}$ be a KY tensor of order $p$ in a maximally symmetric space. Then, inserting (\ref{Riemann-MaxSym}) into (\ref{DDY-KY}), we find that:
$$ \nabla_{a}\, \nabla_b \,\mathcal{A}_{cdE} \eq -\,(p+1)\,\lambda \,g_{a[b}\,\mathcal{A}_{cdE]}  \,. $$
Thus, comparing with (\ref{CCKY}), we conclude that $\nabla_{b} \mathcal{A}_{cdE}$ is a closed CKY tensor of order $(p+1)$. Note that, since $\bl{\mathcal{A}}$ is a KY tensor, the tensor $\nabla_{b} \mathcal{A}_{cdE}$ is totally skew-symmetric, as it should be in order to be called a CKY tensor. So, we have just proved the following:
\begin{theorem}\label{Theo-CCKY}
If $\bl{\mathcal{A}}$ is a Killing-Yano tensor of order $p$ in a maximally symmetric space then its covariant derivative $\mathcal{H}_{bcdE}\equiv\nabla_b \mathcal{A}_{cdE}$ is a closed conformal Killing-Yano tensor of order $(p+1)$. More precisely, we have that
$$ \nabla_{a}\, \mathcal{H}_{bcdE} \eq 2\,g_{a[b}\,J_{cdE]}\; \textrm{ with }\; J_{cdE} \,\equiv\,  -\,\frac{\lambda}{2}\,(p+1)\,\mathcal{A}_{cdE} \,.$$
\end{theorem}
Conversely, in \cite{Stepanov} it has been proved that in a maximally symmetric space of non-zero curvature every closed CKY tensor is the covariant derivative of a KY tensor. It is worth noting that in \cite{Cariglia} some interesting examples of this theorem are constructed by means of Killing spinors. Moreover, in Sasaki spaces there are some important KY forms whose covariant derivatives are closed CKY tensors as well \cite{Kubiznak-Sasaki}. Since the Hodge dual of a closed CKY tensor is a KY tensor, it follows that $\star\bl{\mathcal{H}}$ is a KY tensor of order $(n-p-1)$. Thus, in a maximally symmetric space one can use a KY tensor to generate another KY tensor. Then, one could, in principle, follow the same procedure and use the new KY tensor of order $(n-p-1)$ to generate one more KY tensor of order $p$. However, it turns out that this third KY tensor is, apart from a multiplicative constant, just the tensor $\bl{\mathcal{A}}$, which represents no new symmetry.

Now, let $\bl{Y}$ be any CKY tensor of order $p$, meaning that
\begin{equation}\label{CKY3}
  \nabla_{a}\, Y_{b_1b_2\cdots b_p} \ma \nabla_{b_1}\, Y_{ab_2\cdots b_p} \eq 2\,g_{a[b_1}\,h_{b_2\cdots b_p]} \ma 2\,g_{b_1[a}\,h_{b_2\cdots b_p]} \,
\end{equation}
holds. Then, thanks to (\ref{MaxSym-RicciWeyl}), (\ref{Def.S}) and (\ref{S=RicciY+WeylY}), we conclude that $h_{b_2\cdots b_p}$ is a KY tensor of order $(p-1)$. So, by means of Theorem \ref{Theo-CCKY}, it is possible to construct the following closed CKY tensor of order $p$:
\begin{equation}\label{H-MaxSym}
  H_{b_1b_2\cdots b_p} \,\equiv\, -\, \frac{2}{\lambda\,p}\,\nabla_{b_1}h_{b_2\cdots b_p} \,.
\end{equation}
Since now the initial KY tensor $\bl{h}$ has order $(p-1)$, Theorem \ref{Theo-CCKY} states that the tensor $\bl{H}$ defined in (\ref{H-MaxSym}) obeys the following equation:
\begin{equation}\label{DH-MaxSym}
  \nabla_{a}\, H_{b_1b_2\cdots b_p} \eq 2 \,g_{a[b_1}\,h_{b_2\cdots b_p]} \,.
\end{equation}
Then, using (\ref{CKY3}) and (\ref{DH-MaxSym}) one can easily see that the tensor
$$  A_{b_1b_2\cdots b_p} \,\equiv\, Y_{b_1b_2\cdots b_p} \me   H_{b_1b_2\cdots b_p}  $$
is a KY tensor of order $p$. Writing the latter equation as
$$  \bl{Y} \eq \bl{A} \ma   \bl{H} \,,   $$
we see that $\bl{Y}$ is the sum of a KY tensor and a closed CKY tensor. Since $\bl{Y}$ is an arbitrary CKY tensor, we have proved the following statement.
\begin{theorem}\label{Theo-Decomposition}
In a maximally symmetric space, any conformal Killing-Yano tensor can be decomposed as the sum of a Killing-Yano tensor and a closed conformal Killing-Yano tensor.
\end{theorem}
As the Hodge dual of a closed CKY tensor is a KY tensor, the latter theorem implies that to each CKY tensor of order $p$ in a maximally symmetric manifold are  associated two KY tensors, one of order $p$, namely $\bl{A}$, and the other of order $(n-p)$, namely $\star\bl{H}$.


Finally, let us prove that the decomposition of Theorem \ref{Theo-Decomposition} is unique. Indeed, suppose that the conformal Killing-Yano tensor $\bl{Y}$ admits two decompositions:
$$  \bl{Y} \eq \bl{A} \ma   \bl{H}  \quad \textrm{ and } \quad \bl{Y} \eq \widetilde{\bl{A}} \ma   \widetilde{\bl{H}}  \,. $$
Then, equating both equations we find that
\begin{equation}\label{A-A'}
   \bl{A} \me \widetilde{\bl{A}} \eq \widetilde{\bl{H}} \me \bl{H}  \,.
\end{equation}
But, the left hand side of (\ref{A-A'}) is a KY tensor, while the right hand side is a closed CKY tensor. If a tensor is simultaneously a KY tensor and a closed CKY tensor then it is covariantly constant. Therefore, we conclude that $(\bl{A} - \widetilde{\bl{A}})$ and $(\bl{H} - \widetilde{\bl{H}})$ are both covariantly constant. However, as we shall prove, it turns out that, besides the zero tensor, an $n$-dimensional maximally symmetric space with $\lambda\neq0$ admits no covariantly constant skew-symmetric tensor of rank $p<n$. Indeed, let $F_{a_1\cdots a_p}= F_{[a_1\cdots a_p]}$ be an anti-symmetric covariantly constant tensor,
$$  \nabla_a\, F_{b_1\cdots b_p}  \eq 0 \,.  $$
Then, using the Ricci identity, we have that
$$  0 \eq \lef \nabla_a\,\nabla_b \me \nabla_b\,\nabla_a \rig F_{c_1\cdots c_p} \eq \sum_{i=1}^{p} R_{abc_i}^{\ph{abc_i}e} \, F_{c_1\cdots \check{c}_i\, e \, c_{i+1}\cdots c_p} \,,$$
where the symbol $\check{c}_i$ means that the index $c_i$ is absent. Contracting the latter equation with $g^{ac_1}$ and using (\ref{Riemann-MaxSym}), eventually we are led to
$$  \lambda\, (n-p)\, F_{bc_2\cdots c_p} \eq 0\,.  $$
So, if $p\neq n$ and $\lambda\neq 0$, the covariantly constant skew-symmetric tensor $\bl{F}$ must be the zero tensor. In particular, if the order of the CKY tensor $\bl{Y}$ is less than $n$, we have that $(\bl{A} - \widetilde{\bl{A}})=0$ and $(\bl{H} - \widetilde{\bl{H}})=0$, proving the following result.
\begin{theorem}\label{Theo-Unique}
In a non-flat maximally symmetric space of dimension $n$, the decomposition of a CKY tensor of order $p<n$ as the sum of a KY tensor and a closed CKY tensor is unique. Moreover, apart from the zero tensor, this manifold admits no covariantly constant skew-symmetric tensor of rank $p<n$.
\end{theorem}
Hopefully, the results presented in this section will be of relevance for the study of asymptotic symmetries and gravitational charges in spacetimes that are asymptotically (A)dS \cite{JezierskyAF,Kastor}.

\section{Conclusions}

The significance of the present work relies on unquestionable relevance of symmetries, and its associated conserved charges, in physics. Conservation laws are of special importance to gravitational theories, since the field equations are usually non-linear and, therefore, really hard solve without suitable tools. That is the reason why Killing-Yano and conformal Killing-Yano tensors have proved to be so helpful, since they provide conserved quantities. For instance, the existence of a closed CKY tensor enabled the integration of the geodesic equation as well as Klein-Gordon and Dirac equations on the Kerr-NUT-(A)dS spacetimes of arbitrary dimension \cite{Frolov_KY}. With this physical context in mind, here it has been obtained the integrability conditions for the existence of KY and CKY tensors of arbitrary order in manifolds of arbitrary dimension. Particularly, one of the main contributions of this article have been putting the integrability conditions for the CKY tensors in an explicit conformally invariant form and working out some of their consequences. For instance, it has been proved that, differently to the well-known case of a CKY tensor of order two, the divergence of a CKY tensor of higher order in an Einstein manifold generally is not a KY tensor. Moreover, the analysis of how the existence of a CKY tensor of order two constrains the Petrov classification in 4-dimensional manifolds has been extended to arbitrary signature.

The (C)KY tensors earned the name of hidden symmetries because, generally, it is highly non-trivial to integrate (C)KY equation from scratch. It turns out that the integrability conditions obtained in the present article can be quite helpful for accomplishing this task, as they constrain the algebraic form of the (C)KY tensor which, in turn, eliminates several degrees of freedom in the general ansatz of a CKY tensor. In addition, the integrability conditions presented here can be of relevance for understanding the integrability of Einstein's equation as well. For example, the Petrov classification of a 4-dimensional spacetime admitting a non-reducible KY tensor must be either type $D$ or $O$, where the latter type represents the trivial case of a vanishing Weyl tensor. Since Einstein's vacuum equation can be completely integrated for the type $D$ class of spacetimes \cite{typeD}, there should be a connection between the existence of CKY tensors and the integrability of Einstein's equation. Indeed, it turns out that all type $D$ vacuum solutions admit at least two Killing vectors and one CKY tensor of order two \cite{StephaniBook}. It would be interesting to look for similar relations in higher dimensions.

It is believed that general relativity is just the low-energy effective theory of a quantum theory of gravity. So, at higher energies the Einstein-Hilbert action should be corrected by terms of higher order in the curvature. Besides, there have been several attempts of modifying Einstein's equation in order to explain the issues of dark matter, dark energy, and cosmological inflation, as exemplifies the so-called $f(R)$ theories \cite{f(R)}. As long as the connection used in these theories is metric-compatible and torsion-free, the KY and CKY tensors remain being objects of great relevance. Thus, as a consequence, the integrability conditions obtained here should also be a valuable tool on the study of these modified theories of gravity.\footnote{Note that Einstein's equation have not been assumed to hold throughout this article.} Since gravitational theories with torsion are also of great interest for their richness, specially in the presence of matter fields with intrinsic spin, its is natural to extend the present work to the case of connections with torsion. This is the plan for future researches.

\section*{Acknowledgments}
I want to thank the Brazilian funding agency CAPES (Coordena\c{c}\~{a}o de Aperfei\c{c}oamento de Pessoal de N\'{\i}vel Superior) for the financial support and to my post-Doc supervisor, Professor Bruno Carneiro da Cunha, for the freedom of letting me follow my own research project. I am also grateful to Tsuyoshi Houri for pointing out the existence of Ref. \cite{Kashiwada-Int} as well as for the valuable comments.


\end{document}